\documentclass[onecolumn,
amsmath,amssymb 
]{revtex4}

\usepackage{graphicx}
\usepackage{amsbsy} 
\usepackage{amsmath}
\usepackage{amssymb}
\usepackage{amsfonts}
\usepackage{color}
\usepackage{mathrsfs}
\usepackage{appendix}
\usepackage{color}
\usepackage{mathrsfs}
\usepackage{times,txfonts}
\usepackage{subfigure}

\newtheorem{Thm}{Theorem}
\newtheorem{Lem}[Thm]{Lemma}
\newtheorem{Prop}[Thm]{Proposition}

\newtheorem{Def}[Thm]{Definition}
\newtheorem{Rem}[Thm]{Remark}

\newcommand{\ket}[1]{\left|{#1}\right\rangle}
\newcommand{\bra}[1]{\left\langle{#1}\right|}

\newcommand{\stPhi}{\left| \Phi \right\rangle ^{\otimes 3}}

\begin{document}

\title{Quantum advantage through the magic pentagram problem}


\author{Haesol Han}
\affiliation{Department of Mathematics and Research Institute for Basic Sciences, Kyung Hee University, Seoul 02447, Korea} 

\author{Jeonghyeon Shin}
\affiliation{Department of Mathematics and Research Institute for Basic Sciences, Kyung Hee University, Seoul 02447, Korea} 

\author{Minjin Choi}
\affiliation{Department of Mathematics and Research Institute for Basic Sciences, Kyung Hee University, Seoul 02447, Korea}

\author{Byung Chan Kim}
\affiliation{Department of Mathematics and Research Institute for Basic Sciences, Kyung Hee University, Seoul 02447, Korea} 

\author{Soojoon Lee}
\affiliation{Department of Mathematics and Research Institute for Basic Sciences, Kyung Hee University, Seoul 02447, Korea}

\date{\today}

%


\begin{abstract}
Through the two specific problems, the 2D hidden linear function problem and the 1D magic square problem, Bravyi {\it et al.}
 have recently shown that there exists a separation between $\mathbf{QNC^0}$ and $\mathbf{NC^0}$, 
where $\mathbf{QNC^0}$ and $\mathbf{NC^0}$ are the classes of polynomial-size and constant-depth quantum and classical circuits with bounded fan-in gates, respectively. 
In this paper, we present another problem with the same property, the magic pentagram problem 
based on the magic pentagram game, 
which is a nonlocal game. In other words,  
we show that the problem can be solved with certainty by a $\mathbf{QNC^0}$ circuit 
but not by any $\mathbf{NC^0}$ circuits.
\end{abstract}

\maketitle


\section{Introduction}
\label{Introduction}

Although Shor's factoring algorithm~\cite{Shor97} tells us that 
there exists a quantum algorithm which can be almost exponentially faster 
than any known classical algorithms,   
it has still been unproven that 
there do not exist any classical algorithms 
which can be faster than the Shor's algorithm.
Similarly, quantum speed-up in most studies on quantum algorithms 
has been proved by comparison with the most efficient known classical algorithms 
or computational complexity assumptions.

Recently, it has been rigorously proved~\cite{BGK18,BGKT20} that 
quantum computers can outperform classical computers 
by showing that 
there exists a problem which can be solved on a quantum computer with constant time independent on input size, but cannot be solved on any classical (probabilistic) computers with constant time. 
Furthermore, it has also been shown that such a problem is related with quantum nonlocality, which is one of the unique quantum features with no classical counterpart. 

In particular, Bravyi {\it et al.}~\cite{BGKT20}  have shown that 
a problem called the magic square problem can be defined
by exploiting the magic square game~\cite{Mermin90,Peres90,Mermin93,BBT05} based on quantum nonlocality,  
and the problem provides us with a rigorously provable quantum advantage. 
Interestingly, there is another game similar to the magic square game, which is called the magic pentagram game~\cite{Mermin93,KM17}. 
Note that the magic square game and the magic pentagram game have similar quantum strategies based on the two and three copies of Bell states, respectively.
Hence it is natural to ask a question about whether from the magic pentagram game we can construct the magic pentagram problem which such a
quantum advantage can be derived from even though the magic pentagram game has the more complicated structure than the magic square game. 

In this paper, we give a completely affirmative answer to the question. 
In other words, as in the case of the magic square game, we construct 
the magic pentagram problem related to the magic pentagram game, 
and we prove that the problem can be solved by a $\mathbf{QNC^0}$ circuit but not by any $\mathbf{NC^0}$ circuits, 
where $\mathbf{QNC^0}$ and $\mathbf{NC^0}$ are the classes of polynomial-size and constant-depth quantum and classical circuits with bounded fan-in gates (unbounded fan-out gates are allowed in $\mathbf{NC^0}$ circuits), respectively.

Before getting into our main results, we review the magic pentagram game in the following subsection. 


\subsection{Magic pentagram game}

\begin{figure}[t]
	\centering
	\subfigure[]{
		\includegraphics[width=.3\columnwidth]{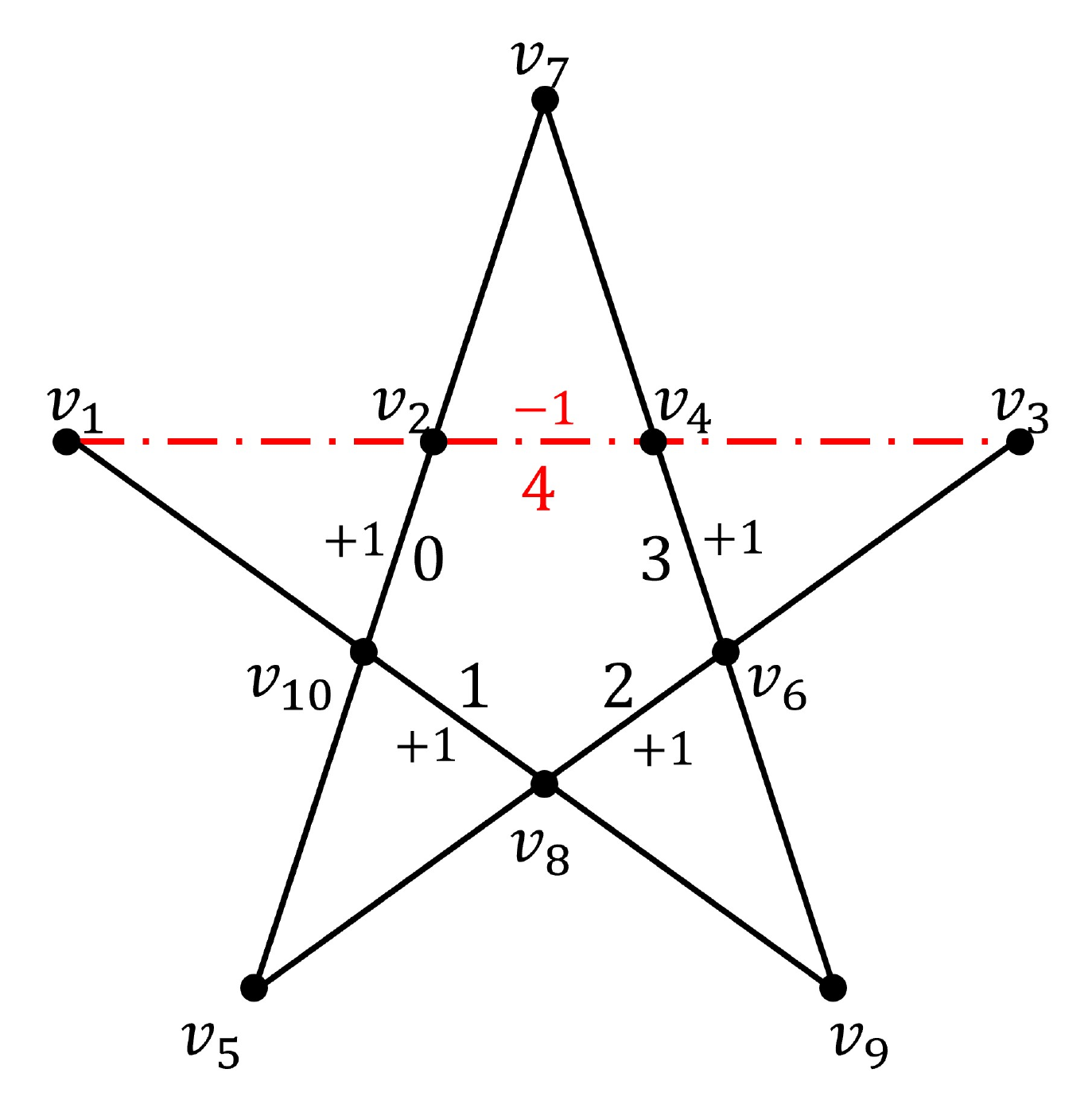}
		\label{MP}
		}
	\hfill
	\subfigure[]{
		\includegraphics[width=.3\columnwidth]{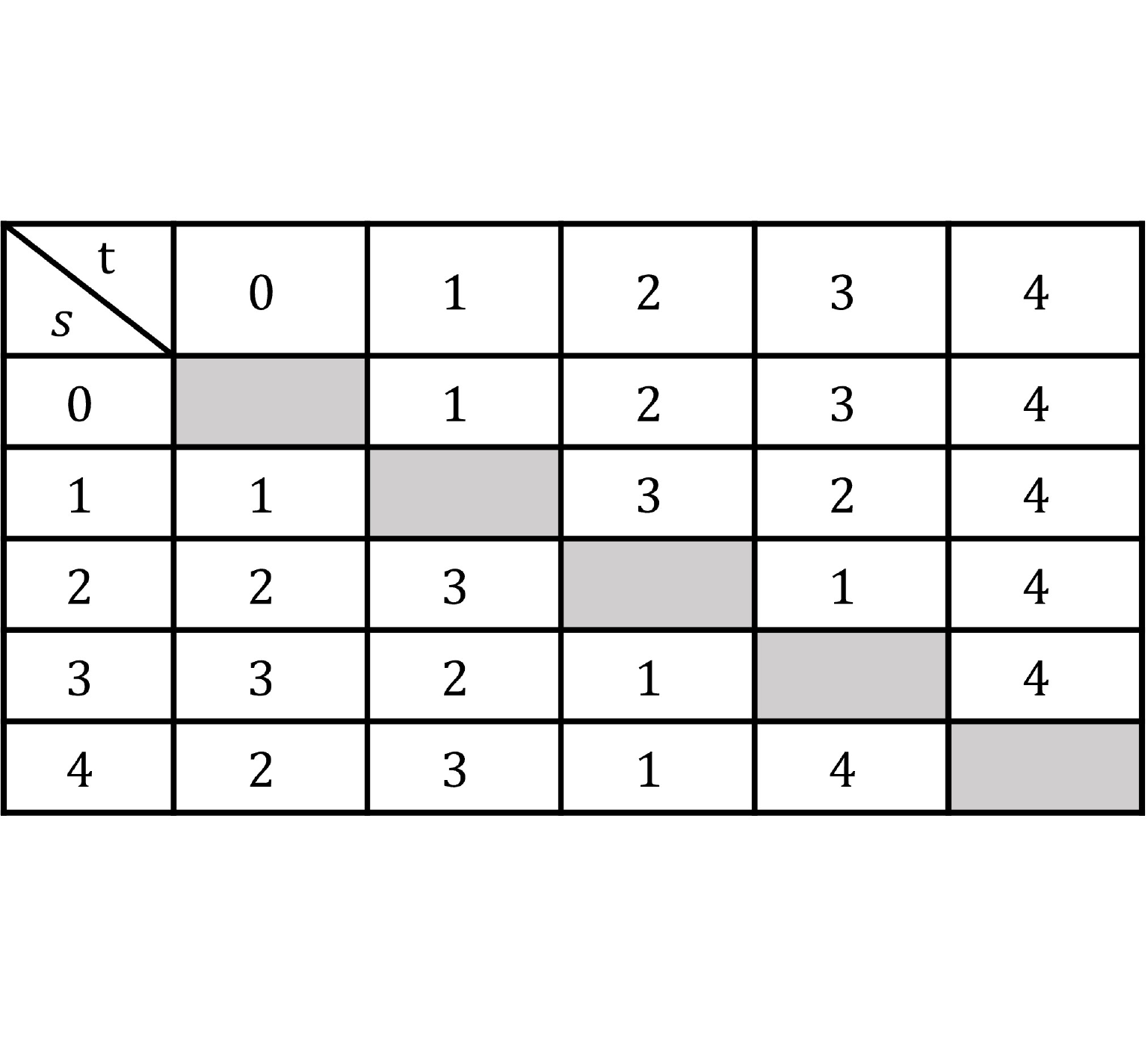}
		\label{funcl}
		}
	\hfill
	\subfigure[]{
		\includegraphics[width=.3\columnwidth]{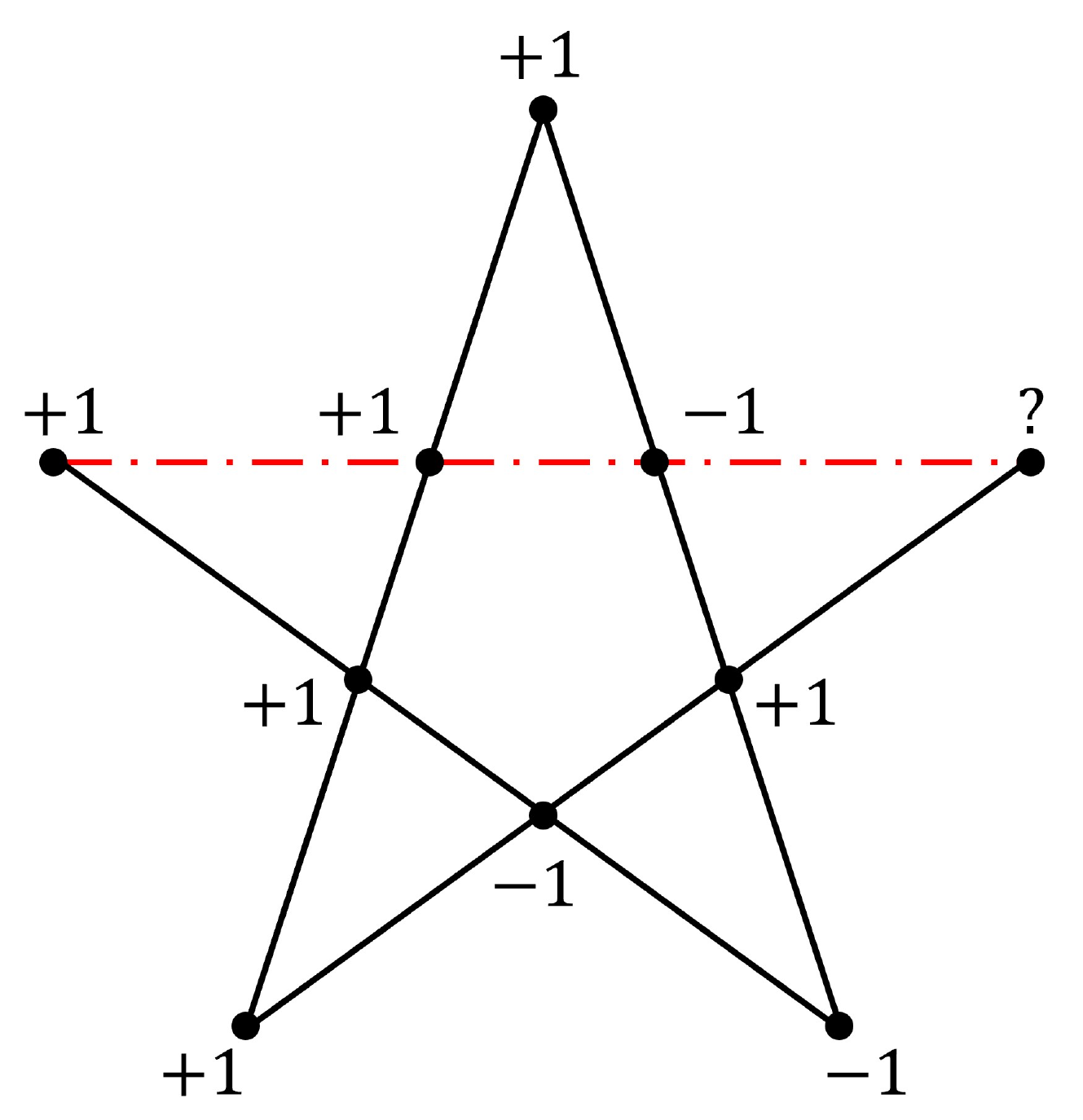}
		\label{strategy}
		}
	\caption{
		(a)~Magic pentagram.
		Each hyperedge corresponds to a number in  $\{0,1,2,3,4 \}$ or $\{000,001,010,011,100 \}$. For each hyperedge $s$, the number $\pm1$ next to $s$ in the figure means the value $e(s)$ in the first winning condition.
		(b)~Definition of $o_s (t)$.
		If a player receives a hyperedge $s$, then his or her $j$-th assignment is given to the vertex at which $s$ intersects with $t$ for some hyperedge $t$ with $o_s(t)=j$. 
		For example, if a player receives a hyperedge $0$ then the player assigns the number $\pm 1$ to the vertices on the hyperedge in the following order: 
		$v_{10}$, $v_{5}$, $v_{7}$ and $v_{2}$ in Figure~\ref{MP}, because $o_0(j)=j$ for $j\in\{1, 2, 3, 4\}$.
		(c)~Example of the classical strategy winning the magic pentagram game. Through this strategy, the players can win the game with probability 19/20.
		}
\end{figure}

The magic pentagram game~\cite{Mermin93,KM17} is a win-win game which requires two players with a referee, that is, either both of the players win or both lose.
The two players, Alice and Bob, start the game with preparing a pentagram like Figure~\ref{MP}.
From the referee, they receive two distinct random numbers $x$ and $y$ in $\{0,1,2,3,4 \}$  respectively, where $x$ and $y$ mean two different hyperedges of the pentagram.
Each player assigns the number $+1$ or $-1$ to each of the four vertices on his or her hyperedge in the following order determined by the function $o_s(t)$ in the table of Figure~\ref{funcl}: 
if a player receives a hyperedge $s$, then his or her $j$-th assignment is given to the vertex at which the hyperedge $s$ intersects with the hyperedge $t$ satisfying $o_s(t)=j$. 
For example, if the hyperedge $0$ is given to a player then the player assigns the numbers $\pm 1$ to the vertices on the hyperedge in the order, $v_{10}$, $v_{5}$, $v_{7}$ and $v_{2}$ 
in Figure~\ref{MP}, since $o_0(j)=j$ for $j\in\{1, 2, 3, 4\}$.
Although the players can discuss before the game, once the game starts, then their communication is not allowed.
We say that the players win the game when the following two winning conditions are satisfied:

\begin{enumerate}
	\item The product of the values assigned to the four vertices on each player's hyperedge $s$ is equal to the value of the following function $e$, 
		where $e$ is the function from the set $\{0, 1, 2, 3, 4\}$ to the set $\{+1,-1\}$ defined by 
		$e(s)=(-1)^{\delta_{s,4}}$, that is, 
		\[
		e(s)=
		\begin{cases}
		+1 &\mathrm{for}~s\neq 4, \\
		-1  &\mathrm{for}~s= 4.
		\end{cases}
		\]
		In other words, let $z=(z^1,z^2,z^3,z^4)$ in $\{ +1, -1 \}^4 $ be an assignment given to the four vertices on the hyperedge $s$ of a player,
		then $\prod_{i=1}^{i=4}z^i =e \left( s \right)$.
	\item If Alice and Bob receive the hyperedges $x$ and $y$,
	          and $z=(z^1,z^2,z^3,z^4)$ and $w=(w^1,w^2,w^3,w^4)$ in $\{ +1, -1 \}^4 $ are their assignments given to 
	          the vertices on the hyperedges,
	          then both of them return the same value on the vertex at which $x$ and $y$ intersect, 
	          that is, $z^{o_{x}(y)}=w^{o_{y}(x)}$.
\end{enumerate}


It is clear that we cannot construct the classical strategy which always makes the players win the magic pentagram game,
since there is no way to assign the numbers $\pm 1$ to all vertices such that
the two winning conditions are satisfied, as we can see through an example in Figure~\ref{strategy}.
 In particular, the players can exchange any amount of classical information at the outset of the game, 
and can employ shared randomness in the classical strategy. 
However, they cannot always win the game, which is precisely shown in the following proposition.

\begin{Prop}\label{MPGclassical}
	The maximal probability that a classical (probabilistic) strategy makes the players win the magic pentagram game is $ 19 / 20 $.
\end{Prop}


On the other hand, it has been known that 
there exists a quantum strategy~\cite{Mermin93,KM17} which always allows the players to win the game with certainty
if they share a proper number of copies of the Bell states $\ket{\Phi} = \frac{1}{\sqrt{2}} \left( \ket{00} + \ket{11} \right)$ in advance, as follows. 

\begin{figure}
	\centering
	\subfigure[]{
		\includegraphics[width=.45\columnwidth]{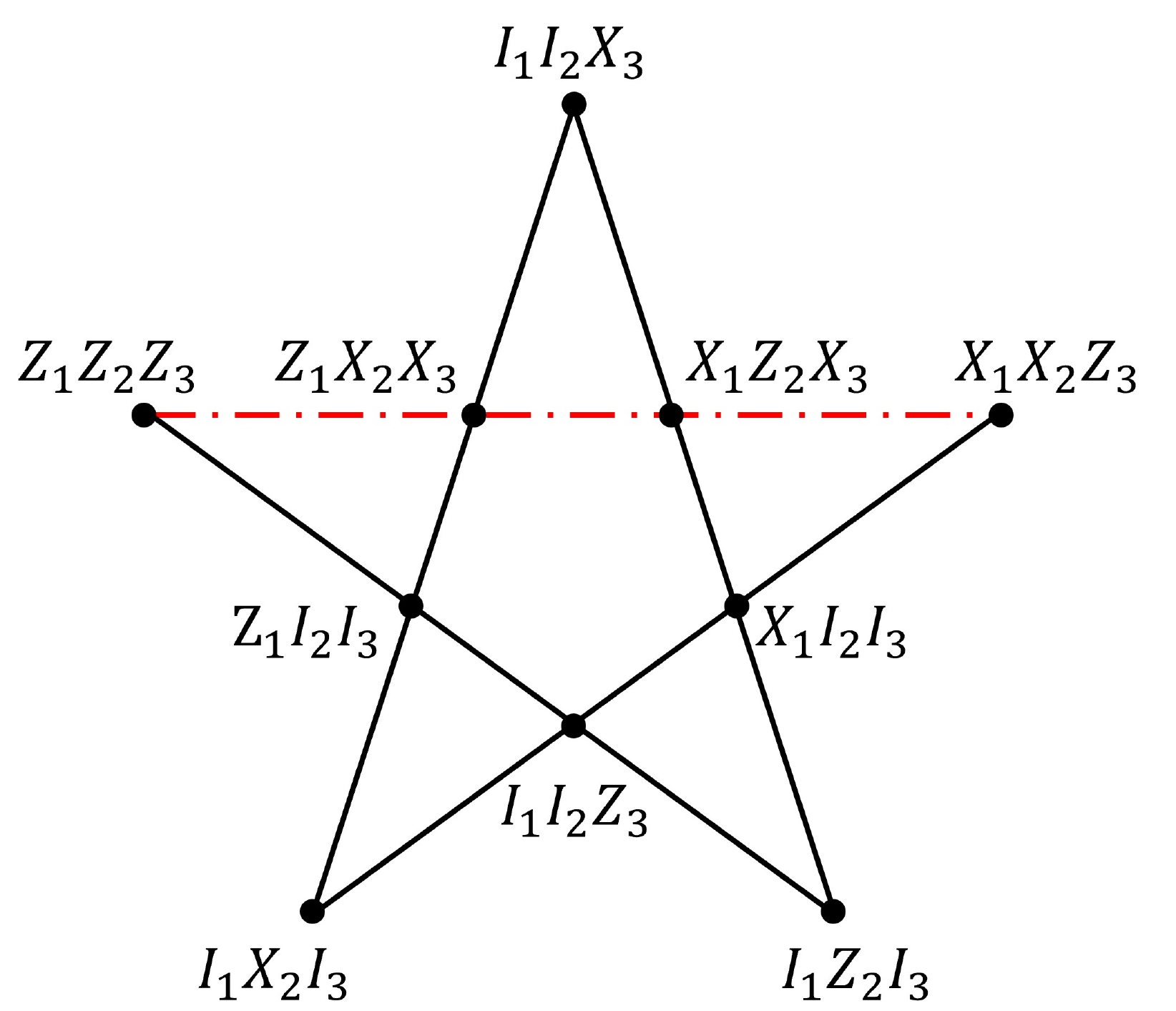}
		\label{Ob}
		}	
	\hfill
	\subfigure[]{
		\includegraphics[width=.45\columnwidth]{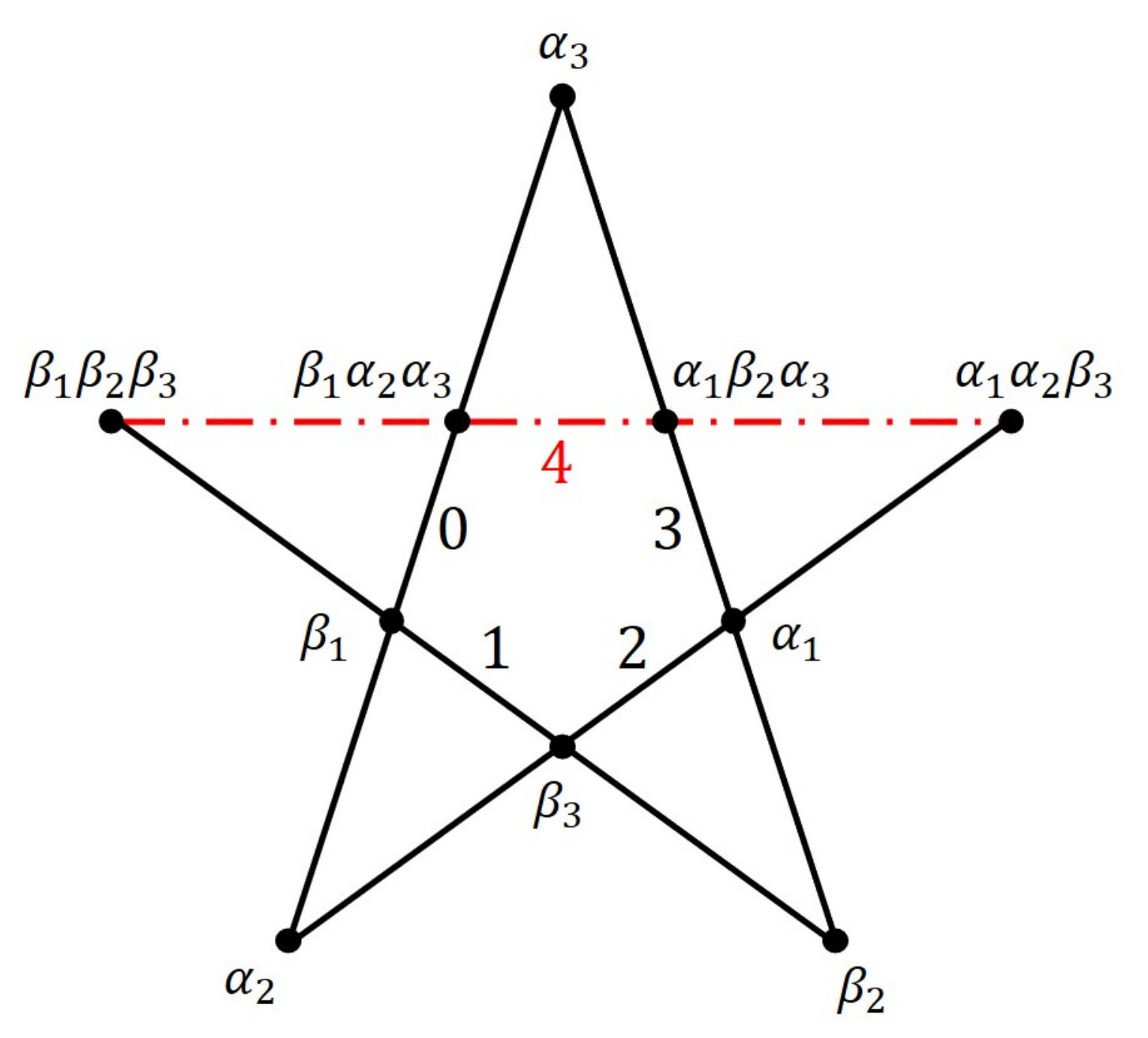}
		\label{defL}
		}
	\caption{(a)~Observables for the magic pentagram game.
	(b)~Definition of $\mathcal{L}_{x,y} 
	\left( {\alpha_1 , \beta_1 , \alpha_2 , \beta_2 , \alpha_3 , \beta_3} \right)$: 
	The function value corresponds to the vertex at which the hyperedges $x$ and $y$ intersect;
	for example, $\mathcal{L}_{0,4} 
	\left( {\alpha_1 , \beta_1 , \alpha_2 , \beta_2 , \alpha_3 , \beta_3} \right)=
	\beta_1\alpha_2\alpha_3$.}
\end{figure}

\begin{Prop}\label{MPGquantum}
	Players can win the magic pentagram game by sharing the maximally entangled state $\stPhi$
	and measuring with the observables corresponding to the vertices as in Figure~\ref{Ob}.
\end{Prop}

Hence the magic pentagram game can be considered as an example to demonstrate the quantum nonlocal characteristics as other nonlocal games
since  the quantum strategy allows the players to share unlimited amounts of entanglement and to perform local quantum operations on their qubits 
while the classical one allows them to share randomness~\cite{BBT05}.

This paper is organized as follows.
In Sec.~\ref{sec:gMPG},
we generalize the magic pentagram game, 
and construct the magic pentagram problem from the generalized game in Sec.~\ref{sec:MPP}. 
In Sec.~\ref{sec:MPP_NC0}, 
we prove that the same quantum advantage as in the magic square problem 
can be gained from the magic pentagram problem. 
Finally, we conclude with discussion on our results in Sec.~~\ref{sec:discussion}.
For the readability, we defer all the proofs to the Appendix.


\section{Generalized magic pentagram game}
\label{sec:gMPG}

In this section, we define the generalized magic pentagram game,
and show that the players can win the game with certainty by an entanglement-based quantum strategy, while the win probability of the classical strategy is at most $19/20$.


For each pair of hyperedges $x$ and $y$ of the pentagram in Figure~\ref{defL},  
let us define a function 
$\mathcal{L}_{x,y}:\{+1, -1\}^6 \rightarrow \{ +1,-1 \}$
by the way that 
$\left( {\alpha_1 , \beta_1 , \alpha_2 , \beta_2 , \alpha_3 , \beta_3} \right)
\in \{+1, -1\}^6$ maps to 
the value corresponding to the vertex at which $x$ intersects with $y$ as in Figure~\ref{defL}. For example, when $x=0$ and $y=4$, $\mathcal{L}_{0,4} 
	\left( {\alpha_1 , \beta_1 , \alpha_2 , \beta_2 , \alpha_3 , \beta_3} \right)=
	\beta_1\alpha_2\alpha_3$.

For each  $\left( {\alpha_1 , \beta_1 , \alpha_2 , \beta_2 , \alpha_3 , \beta_3} \right)\in\{+1,-1\}^6$, 
the {\em generalized magic pentagram game} with 6 parameters 
$\alpha_i$'s and $\beta_i$'s 
is defined as follows. 
The rule of the generalized magic pentagram game is essentially equivalent to that of the original magic pentagram game
except for the second winning condition.
The following two conditions are the winning conditions of the generalized magic pentagram game: 
for players' hyperedges $x$ and $y$, 
 let $z=(z^1,z^2,z^3,z^4)$ and $w=(w^1,w^2,w^3,w^4)$ be their assignments given to the vertices on the hyperedges $x$ and $y$, respectivley.
\begin{enumerate}
	\item $\prod_{i=1}^{i=4}z^i =e \left( x \right)$ and $\prod_{i=1}^{i=4}w^i =e \left( y \right)$.	
	\item $z^{o_{x}(y)} \cdot w^{o_{y}(x)}=\mathcal{L}_{x,y} \left( {\alpha_1 , \beta_1 , \alpha_2 , \beta_2 , \alpha_3 , \beta_3} \right)$.
\end{enumerate}
\begin{Rem}
\label{Rmk:gMPG0}
Since when $\alpha_s = \beta_s = +1$ for all $s \in \{1,2,3 \}$ the generalized magic pentagram game turns into the original magic pentagram game,
the success probability over all possible classical strategies is less than or equal to $19/20$ as in the original one.
%
Moreover, there exists a quantum strategy that allows the players to win the generalized magic pentagram game with probability 1
as in the original magic pentagram game, which we will see in Proposition~\ref{prop:quantum_strategy}.
\end{Rem}

For $\alpha$ and $\beta$ in $\{+1, -1 \}$, let $\ket{ \Phi_{\alpha , \beta} }$ be a maximally entangled state defined as
\begin{equation*}
	\ket{ \Phi_{\alpha , \beta} } = \left( Z^{ \frac{1}{2} (1-\alpha)} X^{ \frac{1}{2} (1-\beta)} \otimes I \right) \ket{\Phi}.
\end{equation*}
Then the following proposition can be obtained.

\begin{Prop}
\label{prop:quantum_strategy}
	Players can win the generalized magic pentagram game with certainty by sharing the state $ \otimes_{s=1}^3 \ket{ \Phi_{\alpha_s , \beta_s} }$
	and measuring with the observables in Figure~\ref{Ob}, which are the same as ones in the original magic pentagram game.
\end{Prop}

We can now construct a constant-depth quantum circuit of the generalized magic pentagram game, 
and obtain the following proposition 
since  $U(\cdot)$ can be shown to be the unitary operator changing the computational basis to the basis related to the observables in Figure~\ref{Ob} 
from its proof, which will be seen in Appendix~\ref{sec:prop4}.

\begin{figure}[t]
	\centering
	\subfigure[]{
		\includegraphics[width=.5\columnwidth]{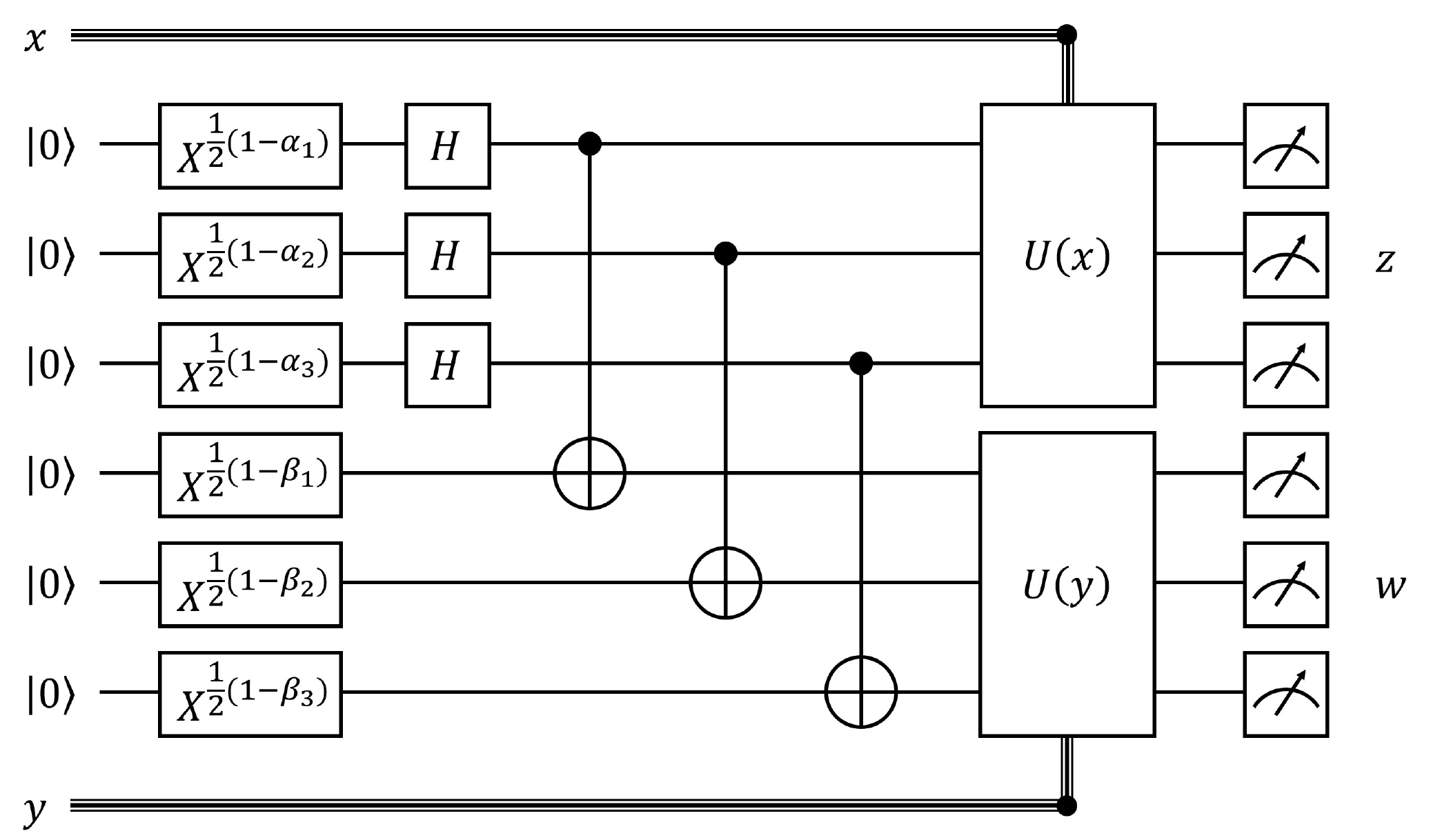}
		\label{MPGcircuit}
		}
	\hfill
	\subfigure[]{
		\includegraphics[width=.4\columnwidth]{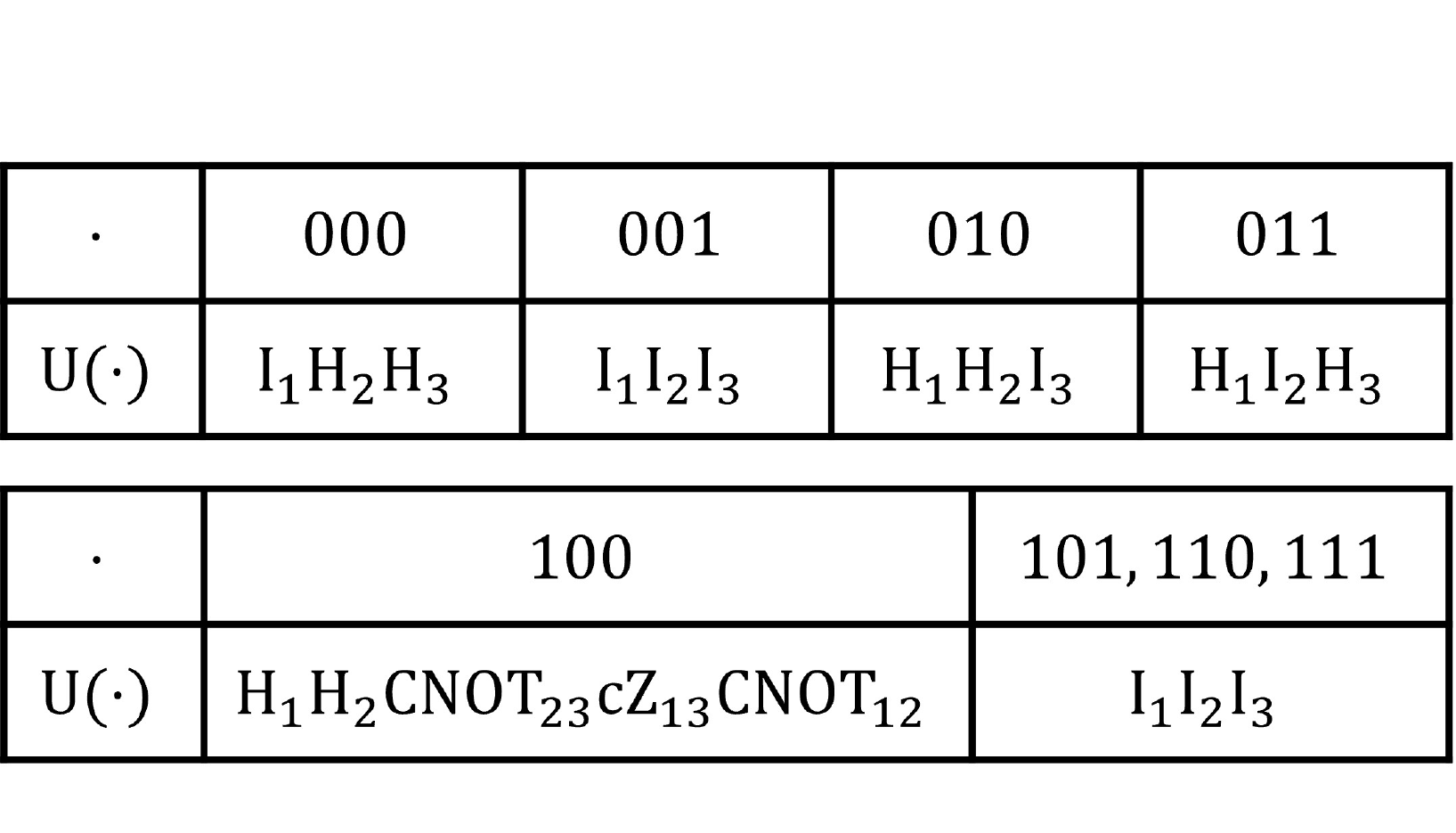} 
		\label{U}
		}
	\caption{(a) Quantum circuit for the generalized magic pentagram game: $x$ and $y$ are three-bit inputs, and $z$ and $w$ are three-bit outputs, 
	which are the measurement outcomes on the final state of the circuit in the computational basis.
	Here, we abuse the notation of $z$ and $w$ to identify 
	the set of three-bit strings with $\{+1,-1\}^3$. 
	Remark that the state before applying $U(x)\otimes U(y)$ in the circuit is  $ \otimes_{s=1}^3 \ket{ \Phi_{\alpha_s , \beta_s} }$ in Proposition~\ref{prop:quantum_strategy}, 
	and $U(\cdot)$ is the unitary operator changing the computational basis to the basis related to the observables in Figure~\ref{Ob}.
	(b) Definition of the gate $U(\cdot)$ in the quantum circuit of Figure~\ref{MPGcircuit}: the values $101,110,111$ are not used in the generalized magic pentagram game, 
	and the gate $U(\cdot)$ represents the basis change associate with the observables in Figure~\ref{Ob}.} 
\end{figure}

\begin{Prop}\label{prop:gMPG}
	The quantum circuit in Figure~\ref{MPGcircuit} 
	exhibits the quantum strategy in Proposition~\ref{prop:quantum_strategy}.
\end{Prop}


\section{Magic pentagram problem}
\label{sec:MPP}

In this section, we define the magic pentagram problem which has $6n$ input bits and $6n$ output bits.
We see that the problem can be solved by a constant-depth quantum circuit with nearest neighbor gates.
On the other hand, we also show that any classical probabilistic circuits composed of bounded fan-in gates cannot solve the problem with certainty.


\begin{figure}
	\centering
	\includegraphics[width=.7\columnwidth ]{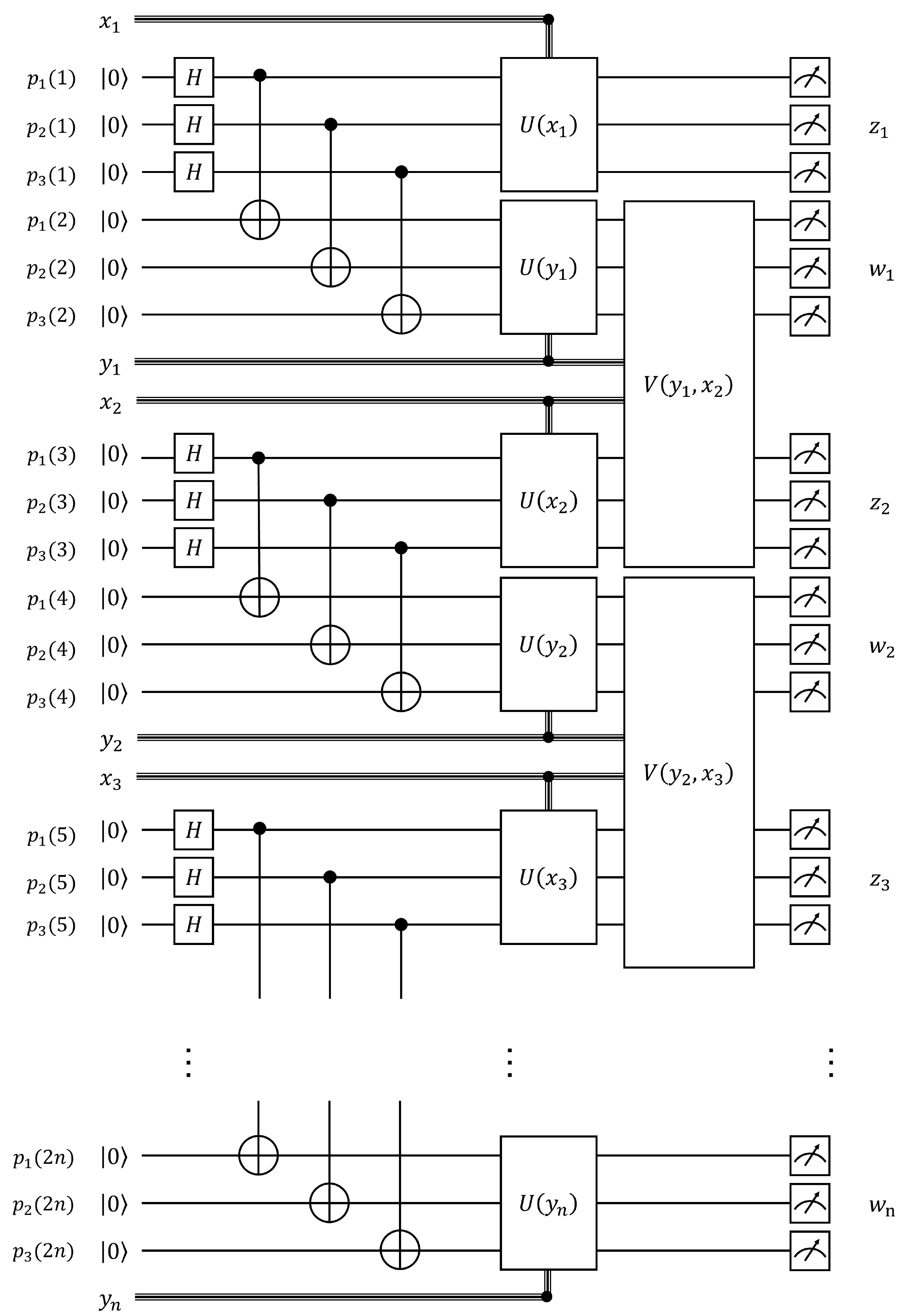}
	\hfil
	\caption{Quantum circuit $\mathcal{C}^{\mathrm{MPP}}$ for the magic pentagram problem.
	In the circuit, there are $6n$ data qubits labelled by $p_s (t)$ with $s \in \{1,2,3 \}$ and $t \in \{1,2, \cdots ,2n \}$. 
	The gate $U$ is the same as in Figure~\ref{MPGcircuit} and  Figure~\ref{U}, 
	and the gate $V$ is defined as in Eq.~(\ref{eq:V}).
	Here, we also abuse the notation of three-bit strings $z_i$ and $w_j$ so that $z_i$ and $w_j$ are in $\{+1,-1\}^3$.
	Remark that for each $j=1,2,\ldots, n$, the partial circuit between classical inputs $x_j$ and $y_j$ before applying the gate $V$ is 
	equivalent to the circuit in Figure~\ref{MPGcircuit} when all $\alpha_i$'s and $\beta_i$'s are one, 
	and $V(y_j,x_{j+1})$ represents the basis change for performing entanglement swapping properly according to the  values $y_j$ and $x_{j+1}$. }
	\label{MPPcircuit}
\end{figure}

In order to define the magic pentagram problem, 
let us consider the quantum circuit $\mathcal{C}^{\mathrm{MPP}}$ in Figure~\ref{MPPcircuit}, 
which is clearly a constant-depth quantum circuit. For $x,y\in\{0,1\}^3$, the gate $V(y,x)$ in Figure~\ref{MPPcircuit} is defined as
\begin{equation}
V(y,x)=
\begin{cases}
    M_{14} M_{25} M_{36}\left[U(y)^{\dagger}\otimes U(x)^{\dagger}\right] & 
    \mathrm{if}~x,y \in \{ 101,110,111 \}~\mathrm{or}~x=y, \\
    I^{\otimes6} & \mathrm{otherwise}
\end{cases}
\label{eq:V}
\end{equation}
where $M_{ij}=(H_i \otimes I_j) \mbox{CNOT}_{ij}$ is the Bell basis change on the $i$-th and $j$-th qubits, mapping the Bell basis to the computational one.
\begin{Rem}
Let $X^{in}$ and $Z^{out}$ be the classical input and the classical output of the circuit $\mathcal{C}^{\mathrm{MPP}}$. 
Then the input $X^{in}$ can be expressed as $X^{in}=(x_1, x_2, \cdots, x_n, y_1, y_2, \cdots, y_n)  \in \{ {0,1} \}^{6n}$ where $x_j , y_j \in \{0,1\}^3$ for $j \in \{1,2, \cdots ,n \}$,
as seen in Figure~\ref{MPPcircuit}.
Assume that $x_k, y_l\in\{000,001,010,011,100\}$
for some $k$ and $l$ with $1\le k<l\le n$, 
and $x_i, y_j\in \{101,110,111\}$ for all $i\neq k$ and $j\neq l$. Then
the two outputs $z_{k}$ and $w_{l}$ after running the circuit $\mathcal{C}^{\mathrm{MPP}}$ is equivalent to 
the outputs $z$ and $w$ after running the circuit in Figure~\ref{MPGcircuit} for some proper $\alpha_i$'s and $\beta_i$'s, 
since entanglement swapping is implemented through the Bell measurements on the qubits $p_s(2t)$ and $p_s(2t+1)$ for $s=1, 2, 3$ and $t=k, k+1,  \ldots, l-1$ in the circuit.
\end{Rem}

We now define the magic pentagram problem by exploiting the input-output relation obtained from the circuit.

\begin{Def}[Magic pentagram problem]
	We say that a circuit with $6n$-bit input and $6n$-bit output solves the {\em magic pentagram problem} if for every input $ X^{in} \in \{ {0,1} \}^{6n}$,
	the circuit outputs $ Z^{out} \in \{ {0,1} \}^{6n}$ such that
	\begin{equation*}
		Pr_{X^{in}} \left[ Z^{out} \right] 
		= { \left| { \bra{ Z^{out} } \mathcal{C}^{\mathrm{MPP}} \left( X^{in} \right) \ket{ 0^{6n}} } \right| }^2 >0,
	\end{equation*}
	where $\mathcal{C}^{\mathrm{MPP}} \left( X^{in} \right)$ is the quantum circuit $\mathcal{C}^{\mathrm{MPP}}$  with the input $X^{in}$ in Figure~\ref{MPPcircuit}.
\end{Def}
Then we can directly have the following theorem by definition of the magic pentagram problem.

\begin{Thm}[Magic pentagram problem is in $\mathbf{QNC}^0$]
	The magic pentagram problem can be solved with certainty by a $\mathbf{QNC}^0$ circuit.
\end{Thm}


\section{Magic pentagram problem is not in $\mathbf{NC^0}$}
\label{sec:MPP_NC0}

In this section, we first 
consider a specific subset of the full instance set for the magic pentagram problem, and then
show that  any $\mathbf{NC^0}$ circuits with the full instance set as input, cannot solve the magic pentagram problem with probability greater than $19/20$ 
for a randomly chosen input from the subset.  

For each $ 1\le k<l\le n$, let $S_{k,l}$ be the set of all $6n$-bit strings $(x_1, x_2, \cdots, x_n, y_1, y_2, \cdots, y_n)$ 
such that $x_k, y_l\in \{000,001,010,011,100\}$,  $x_i=111$ for all $i\neq k$, and $y_{j}=111$ for all $j\neq l$.
and let $S$ be an instance subset of the magic pentagram problem defined as 
\[
S=\bigcup_{k<l}S_{k,l}.
\]

\begin{Rem}
\label{Rmk:gMPG}
Assume that $X^{in}=(x_1, x_2, \cdots, x_n, y_1, y_2, \cdots, y_n)$ is randomly chosen from $S$. Then there exist $1\le k<l\le n$ such that
$x_i=111$ and $y_j=111$ for all $i\neq k$ and $j\neq l$
, 
and $x_k$ and $y_l$ are random numbers in $\{000,001,010,011,100\}$. 
Let  
\[
Z^{out}=(z_1, z_2, \cdots, z_n, w_1, w_2, \cdots, w_n) \in \{ {+1,-1} \}^{6n}
\]
 be a measurement outcome after applying the circuit $\mathcal{C}^{\mathrm{MPP}}\left( X^{in} \right)$ to $\ket{0^{6n}}$, 
and for $i \in \{ 1, 2, 3 \}$, let 
	\begin{equation*}
		\alpha_{i}=\prod_{j=k}^{l-1} w_j^i  
		\mbox{\quad and \quad}  
		\beta_{i}=\prod_{j=k}^{l-1} z_{j+1}^i,
	\end{equation*}
where $z_{j+1}=z_{j+1}^1z_{j+1}^2z_{j+1}^3$	
and $w_{j}=w_{j}^1w_{j}^2w_{j}^3$ are in $\{+1,-1\}^3$. 
Then by employing the way almost same as in Lemma~3 of the Bravyi {\it et al.}'s result on the magic square problem~\cite{BGKT20}, 
we can see that for players' hyperedges $x_k$ and $y_l$, 
the two quadruples, 
 $z=\left(z_k^1,z_k^2,z_k^3,z_k^1z_k^2z_k^2e(x_k)\right)$ and $w=\left(w_l^1,w_l^2,w_l^3,w_l^1w_l^2w_l^3e(y_l)\right)$, 
 satisfy the second winning condition of the generalized magic pentagram game with 6 parameters $\alpha_i$'s and $\beta_j$'s, that is, 
\[
z_k^{o_{x_k}(y_l)} \cdot w_l^{o_{y_l}(x_k)}
=\mathcal{L}_{x_k,y_l} \left( {\alpha_1 , \beta_1 , \alpha_2 , \beta_2 , \alpha_3 , \beta_3} \right).
\]
But, this does not imply that 
the players win the generalized magic pentagram game, 
since it is not guaranteed that 
$z$ and $w$ are independent of $y_l$ and $x_k$, respectively.
\end{Rem}

Applying the properties in the Appendix~\ref{sec:lightcones} to the magic pentagram problem, we can derive our main theorem described below.

\begin{Thm}[Magic pentagram problem is not in $\mathbf{NC}^0$]
        \label{MPP_NC0} 
	Let $\mathcal{C}$ be
	a classical probabilistic circuit with $6n$-bit inputs, $6n$-bit outputs and gates of fan-in at most $B$, and 
	assume that $\mathcal{C}$ solves the magic pentagram problem with probability $p>19/20$ for any arbitrary random input from $S$.
	Then the depth of $\mathcal{C}$ is at least
	\begin{equation}
		\frac{1}{2}{\log_{B} \left[\frac{n}{216} \left(p-\frac{19}{20} \right) \right]}.
		\label{eq:depth_bd}
	\end{equation}
\end{Thm}

The value in Eq.~(\ref{eq:depth_bd}) cannot be bounded above by a constant number, since as $n$ tends to infinity, it also goes to infinity. 
Therefore, Theorem~\ref{MPP_NC0} implies that the magic pentagram problem cannot be solved with certainty by any $\mathbf{NC}^0$ circuits.

\begin{Rem}
In Ref.~\cite{BGKT20}, there is a similar result to Theorem~\ref{MPP_NC0}, 
which states that if a classical circuit with gates of fan-in at most $B$ solves the magic square problem with probability at least $9/10$ for any arbitrary random input from a proper subset, 
then the depth of the circuit has a lower bound 
\begin{equation}
		\frac{1}{2}{\log_{B} \left( 0.00001 n\right)}.
		\label{eq:depth_bd_msp}
\end{equation}
As in Theorem~\ref{MPP_NC0}, the above result in Ref.~\cite{BGKT20} can slightly be improved as follows: 
if a classical circuit with gates of fan-in at most $B$ solves the magic square problem with probability $p>8/9$ for any arbitrary random input from the subset, 
then the depth of the circuit is at least
\begin{equation}
		\frac{1}{2}{\log_{B} \left[\frac{n}{80} \left(p-\frac{8}{9} \right) \right]},
		\label{eq:depth_bd_msp2}
\end{equation}
which is greater than the original lower bound of the circuit depth in Eq.~(\ref{eq:depth_bd_msp}) for $p\ge 0.89$.
\end{Rem}


\section{Conclusion and discussion}
\label{sec:discussion}

We have first considered the magic pentagram game which is based on quantum nonlocality, and
have constructed the magic pentagram problem by exploiting a quantum strategy to win the game, 
and have then shown that the problem can be solved with certainty by a $\mathbf{QNC^0}$ circuit, 
whereas no $\mathbf{NC^0}$ circuits can solve the problem with certainty. 
Hence, we can conclude that 
the magic pentagram problem presented here is another example 
to show the explicit separation between shallow quantum circuits and bounded fan-in shallow classical ones. 

We note that there exists a problem to solve with near certainty using a noisy shallow quantum circuit 
if the noise rate is below a certain threshold value 
while the problem cannot be solved with high probability by any noise-free shallow classical circuits~\cite{BGKT20}. 
The problem is the noise-tolerant version of the magic square problem, which results from the rigidity of the magic square game~\cite{WBMS16}. 
This implies that all near-optimal strategies for the game are approximately equivalent to a unique quantum strategy exploiting quantum entanglement.
Since it was also proved that  the magic pentagram game is rigid~\cite{KM17}, 
if the win probability of a strategy for the magic pentagram game is close to one, then 
the strategy is approximately equivalent to the quantum strategy  
based on three copies of the Bell states presented in Proposition~\ref{MPGquantum}. 
Hence, we may have the same result by defining the noise-tolerant version of the magic pentagram problem 
which can be satisfied with probability close to one by the input-output statistics of a noisy shallow quantum circuit 
as in the result~\cite{BGKT20}. 

In the result of Watts {\it et al.}~\cite{WKST19}, it has been shown that the 2D hidden linear function problem providing the same quantum advantage~\cite{BGK18} as in the magic pentagram problem cannot be solved with certainty even by any $\mathbf{AC^0}$ circuits, 
where $\mathbf{AC^0}$ is the class of polynomial-size and constant-depth classical circuits in which {\em unbounded} fan-in gates are allowed to use. It would be an interesting future work to investigate whether the magic pentagram problem is not in $\mathbf{AC^0}$.

There are several quantum nonlocal games, called the quantum pseudo-telepathy~\cite{BBT05}, 
in which the win probability of quantum players dealing with quantum entanglement 
is even greater than that of classical players with shared randomness but no shared entanglement. 
Therefore, by properly exploiting those games, we could construct various kinds of problems to attain quantum advantage.


\begin{acknowledgements}
This research was supported by the National Research Foundation of Korea (NRF) grant funded
by the Ministry of Science and ICT (MSIT) (Grants No. NRF-2020M3E4A1079678).
S.L. acknowledges support from the MSIT, Korea, 
under the Information Technology Research Center support program (Grant No. IITP-2022-2018-0-01402) 
supervised by the Institute for Information and Communications Technology Planning and Evaluation, 
and the Quantum Information Science and Technologies program of the NRF 
funded by the MSIT (Grant No. 2020M3H3A1105796).
\end{acknowledgements}

\begin{appendices}

\section*{Appendix: Proofs of our results }

The proofs in this paper are essentially based on the ones in Bravyi {\it et al.}'s result on the magic square problem~\cite{BGKT20}, but their details are rather different, since the magic pentagram problem is more complicated than the magic square problem.

\section{Proof of Proposition~\ref{MPGclassical}}
\label{sec:Prop1}

        We here prove this proposition by using our notation, even though the proof has been well-known~\cite{Mermin93,KM17}.
	Assume that there exists a classical winning strategy with $z_{v_j}\in\{+1,-1\}$ for each vertex $v_j$.
	Then multiplying the output values on all hyperedges, we get $\prod_{x=0}^4 e(x)=-1$.
	However, since every pair of hyperedges intersect at exactly only one vertex, the value also becomes $\left(\prod_j z_{v_j}\right)^2$, which must be positive. This leads to a contradiction.
	Thus there do not exist any classical winning strategies to definitely win the magic pentagram game.
	Hence the maximum success probability over all possible probabilistic strategies is less than or equal to $19/20$.
	Moreover, we can easily obtain the classical strategy to win the game with probability $19/20$ as in Figure~\ref{strategy}.

\section{Proof of Proposition~\ref{MPGquantum}}
\label{sec:prop2}

        The  proof of this proposition has already been known~\cite{Mermin93,KM17},
        but we rewrite the proof with our notation. 
	As seen in Figure~\ref{Ob}, any pair of the observables corresponding to vertices on each hyperedge 
	commute with each other.
	Since Alice and Bob share the maximally entangled state $\stPhi$, they get the same outcome from the same observable.
	So, the second winning condition is satisfied.

	If we multiply all four observables on a hyperedge $s$, then we get the identity operator with sign depending on $e(s)$.
	Thus the product of the measurement outcomes becomes $e \left( s \right)$.
	Therefore, the measurement outcomes satisfy the first winning condition. 
	This implies that the players win the magic pentagram game with certainty.

\section{Proof of Proposition~\ref{prop:quantum_strategy}}
\label{sec:prop3}

	First, note that $\ket{\Phi_{\alpha , \beta}}$ can be rewritten as
	\begin{equation*}
		\ket{\Phi_{\alpha , \beta} } = \frac{1}{\sqrt{2}} 
		\left(  (-1)^{\frac{ (1-\alpha)(1-\beta)}{4}} \ket{\frac{1-\beta}{2} , 0} +  (-1)^{\frac{(1-\alpha)(1+\beta)}{4} } \ket{\frac{1+\beta}{2} , 1} \right).
	\end{equation*}
	Thus, if Alice and Bob measure $\ket{\Phi_{\alpha , \beta} }$ by the Pauli matrix $Z$,
	then the outcome is either $\left( \beta , 1 \right)$ or $\left( -\beta, -1 \right)$ 
	since $(-1)^{\frac{1\mp\beta}{2}}=\pm\beta$ for $\beta\in\{+1,-1\}$, and
	the product of the two outcomes hence becomes $\beta$ at all times.
	We also note that $\ket{\Phi_{\alpha , \beta}}$ can be expressed as
	\begin{equation}
		\ket{\Phi_{\alpha , \beta} } = \frac{1}{\sqrt{2}}
		\left( \ket{\frac{1-\alpha}{2}_x , 0_x} +  (-1)^{\frac{1-\beta}{2}} \ket{\frac{1+\alpha}{2}_x , 1_x} \right), 
		\label{bellab}
	\end{equation}
	where $\ket{k_x}=\frac{1}{\sqrt{2}}\left(\ket{0}+(-1)^k\ket{1}\right)$ is the $(-1)^k$-eigenstate of the Pauli matix $X$. 
	Thus, if Alice and Bob measure $\ket{\Phi_{\alpha , \beta} }$ by the Pauli matrix $X$, then the outcome is either $\left( \alpha , 1 \right)$ or $\left( -\alpha , -1 \right)$, 
	and hence  the product of the two outcomes becomes $\alpha$ regardless of the outcome values.
	
	Comparing the observables in Figure \ref{Ob} and the definition of $\mathcal{L}_{x,y}$ in Figure \ref{defL},
	we can see that $\alpha_j$ and $\beta_k$ correspond to $X_j$ and $Z_k$, respectively.
	Consequently, the outcomes satisfy the winning conditions of the generalized magic square game.

\section{Proof of Proposition~\ref{prop:gMPG}}
\label{sec:prop4}

\begin{figure}[t]
	\centering
	\includegraphics[width=.4\columnwidth]{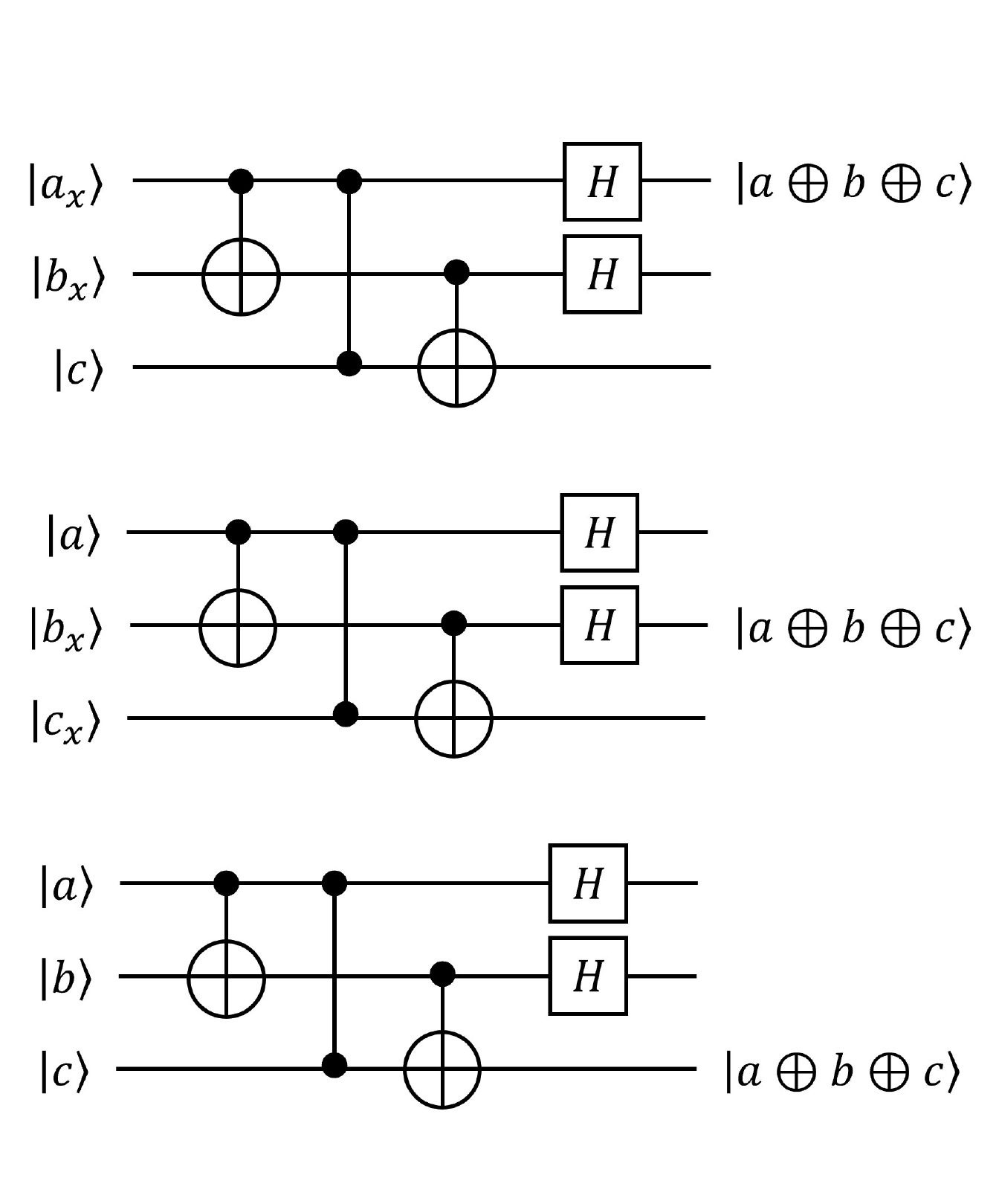}
	\caption{The role of the gate $U(100)$ in Figure~\ref{U}}. 
	\label{Ugate}
\end{figure}

	It can readily be seen that 
	$\ket{ \Phi_{\alpha , \beta} } = \mathrm{CNOT}(H\otimes I)
	\ket{\frac{1-\alpha}{2},\frac{1-\beta}{2}}$,
	and $U(\cdot)$ implements the basis changes necessary for measuring by the observables in Figure~\ref{Ob}. 
	In particular, we can see the role of the gate $U(100)$ from its properties in Figure~\ref{Ugate}.
	The players can assign the outputs $z=(z^1, z^2, z^3) \in \{+1,-1\}^3$ and $w=(w^1,w^2,w^3)\in \{+1,-1\}^3$
	to vertices on the received hyperedges $s$ and $t$, respectively, following the order given from $o_s(t)$.
	In addition, $z^4=z^1 z^2 z^3 e(x)$ and $w^4=w^1 w^2 w^3 e(y)$ can be assigned to the remaining final vertices, respectively.
	Therefore, this completes the proof.

\section{Classical circuits and disjoint lightcones}
\label{sec:lightcones}

In this section, by means of the concept of the lightcones, we first consider the situation that the input values and the output values related to the generalized magic pentagram game as in Remark~\ref{Rmk:gMPG} are independent, 
and then investigate the relation with the magic pentagram problem,   
which is similar to the Bravyi {\it et al.}'s result about the magic square problem~\cite{BGKT20}.

\begin{Def}
    For variables $x_i , z_j \in \{ 0,1 \}$ with $1 \le i\le N$ and $1\le j\le M$ which mean the $i$-th input bit and the $j$-th output bit of a classical circuit $\mathcal{C}$ with $N$ input bits and $M$ output bits, respectively, we say that $\left( x_i , z_j \right)$ are \emph{correlated} if there exists an input string $X^{in} \in \{ 0,1 \}^{N}$ such that $j$-th bit of $\mathcal{C} ( X^{in} )$ changes when the $i$-th bit of $X^{in}$ flips.
	For an input variable $x_i$, let ${L_\mathcal{C}} \left( x_i \right)$ be the set of output bits $z_t$ such that $\left( x_i , z_t \right)$ are correlated,
	which is called the {\em  
	lightcone} of $x_i$. 
	For a set of input bits $I$, 
	we define ${L_\mathcal{C}} \left( I \right)$ 
	by
	\begin{equation*}
		{L_\mathcal{C}} \left( I \right) = \bigcup_{x \in I} {L_\mathcal{C}} \left( x \right).
	\end{equation*}
\end{Def}

We hereafter assume that $\mathcal{C}$ is a depth-$D$ classical probabilistic circuit composed of gates of fan-in at most $B$ 
which has inputs $\left( x_1 , x_2 , \cdots , x_n , y_1 , y_2 , \cdots , y_n \right) \in \{0,1\}^{6n}$
and outputs $\left( z_1 , z_2 , \cdots , z_n , w_1 , w_2 , \cdots , w_n \right) \in \{0,1\}^{6n}$.
For each $ 1 \le k < l \le n $, let $E_{k,l}$ be the subset of $S_{k,l}$ such that 
    \begin{equation*}
		{L_\mathcal{C}} (x_k ) \cap {L_\mathcal{C}} (y_l ) = \emptyset \mbox{\, , \,} 
		{w_l} \notin {L_\mathcal{C}} (x_k )
		\mbox{\, and \,} 
		z_k \notin {L_\mathcal{C}} (y_l ),
	\end{equation*}
and let $E\subseteq S$ be the event defined as $E=\bigcup_{k<l}E_{k,l}$. 
Then we can obtain the following proposition which is almost the same as Lemma~7 in Supplementary Information of the Bravyi {\it et al.}'s result~\cite{BGKT20}.

\begin{Prop}\label{lem1}
	If we choose an input from $S$ uniformly, the probability that $E$ occurs is at least $1 - \frac{216}{n} B^{2D} $.
\end{Prop}

We now assume that a classical circuit $\mathcal{C}$ solves the magic pentagram problem for a randomly chosen input from $S$. Then its output is equal to a measurement outcome on the resulting state after applying the circuit $\mathcal{C}^{\mathrm{MPP}}$ with the input to the initial state $\ket{0^{6n}}$. 
Thus Remark~\ref{Rmk:gMPG} tells us that 
 two quadruples satisfying the second winning condition for the generalized magic pentagram game with some proper 6 parameters can be obtained from the output.
If the event $E$ occurs for a randomly chosen input from $S$ and the input is in $S_{k,l}$ for some $1\le k < l \le n$ then the outputs correlated with $x_k$ are all independent of 
the outputs correlated with $y_l$, 
and $x_k$ and $y_l$ are independent of $w_l$ and $z_k$, respectively. 
Hence, we can show that 
the players with the hyperedges $x_k$ and $y_l$ win the generalized magic pentagram game with the 6 parameters obtained from the output of the circuit $\mathcal{C}$ as in Remark~\ref{Rmk:gMPG}. 
Accordingly, by Remark~\ref{Rmk:gMPG0}, we clearly obtain the following lemma. 

\begin{Lem}\label{lem2}
    Assume that the event $E$ occurs for a randomly chosen input from $S$.
	Then the average probability that $\mathcal{C}$ solves the magic pentagram problem is at most $19/20$.	
\end{Lem}

\section{Proof of Theorem~\ref{MPP_NC0}}
\label{sec:thm3}

	Let $D$ be the depth of $\mathcal{C}$. 
	Then by Lemma~\ref{lem2} and Proposition~\ref{lem1}, we can find the upper bound on $p$ as follows. 
	\begin{align*}
			p=\Pr \left[ \mathcal{C} \mbox{ succeeds} \right]
			&=\Pr\left[E\right]\Pr \left[ \mathcal{C} \mbox{ succeeds} | E \right]
			+ \Pr\left[E^c\right]\Pr \left[ \mathcal{C} \mbox{ succeeds} | E^c \right]\\
			&\le  \Pr \left[ \mathcal{C} \mbox{ succeeds} | E \right] + \left(1 - \Pr \left[ E \right] \right)
			\\&\le \frac{19}{20} + \frac{216B^{2D}}{n},
	\end{align*}
    which implies 
	\begin{equation*}
		B^{2D} \ge \frac{n}{216} \left(p-\frac{19}{20} \right).
	\end{equation*}
	This completes the proof. 

\end{appendices}


\bibliographystyle{spphys}

\end{document}